\documentclass[aps,pre,twocolumn,nofootinbib,floatfix]{revtex4}
\usepackage{amsmath}
\usepackage{epsfig}

\input{tcilatex}

\begin{document}

\title{Effect of electron-phonon interaction on the impurity binding energy
in a quantum wire}
\author{Yueh-Nan Chen}
\email{ynchen.ep87g@nctu.edu.tw}
\author{Der-San Chuu}
\email{dschuu@cc.nctu.edu.tw}
\author{Yuh-Kae Lin}
\affiliation{Department of Electrophysics, National Chiao-Tung University, Hsinchu 30050,
Taiwan}
\date{\today }

\begin{abstract}
The effect of electron-optical phonon interaction on the hydrogenic impurity
binding energy in a cylindrical quantum wire is studied. By using Landau and
Pekar variational method, the hamiltonian is separated into two parts which
contain phonon variable and electron variable respectively. A
perturbative-variational technique is then employed to construct the trial
wavefunction for the electron part. The effect of confined electron-optical
phonon interaction on the binding energies of the ground state and an
excited state are calculated as a function of wire radius. Both the
electron-bulk optical phonon and electron-surface optical phonon coupling
are considered. It is found that the energy corrections of the polaron
effects on the impurity binding energies increase rapidliy as the wire
radius is shrunk, and the bulk type optical phonon plays the dominant role
for the polaron effects.

\end{abstract}

\maketitle

\address{Department of Electrophysics, National Chiao Tung University,
Hsinchu 30050, Taiwan}

\address{Department of Electrophysics, National Chiao Tung University,
Hsinchu 30050, Taiwan}

\address{Department of Electrophysics, National Chiao Tung University,
Hsinchu 30050, Taiwan}

\address{Department of Electrophysics, National Chiao Tung University,
Hsinchu 300, Taiwan}





\section{Introduction}

During the past decades the development of the epitaxial crystal growth
techniques such as molecular beam epitaxy and metal-organic chemical vapor
deposition has made the growth of the quasi-two-dimensional (quantum well)
or quasi-one-dimensional (quantum wire)\cite{1,2,3,4} systems with
controllable well thickness or wire radius became possible. These quantum
structures have been applied to many semiconductor devices, such as
high-electron-mobility transistors. Recent progresses in growth and
fabrication techniques have been able to fabricate the quantum wires with
radii less than 100 $\overset{\circ }{\text{A}}$. Theoretically, the
electronic properties of a hydrogenic impurity in the quantum well\cite%
{5,6,7,8} and the quantum wire \cite{9,10,11,12,13,14} have been studied by
many authors. The impurity binding energies of a quantum wire with infinite
or finite potential barrier \cite{9} and with different shapes of the
cross-section\cite{10,11} have been discussed. The effect of location\cite%
{10,11} of impurities with respect to the wire axis was also studied
previously. The emission line for quantum wires was observed\cite{15} to be
two to three times broader than that of quantum wells and with 6-10 meV
higher binding energy. It is expected that the same properties in quantum
wells were further improved by the reduction of dimensionality to
quasi-one-dimensional quantum wires.\newline

The physics of impurity states in quantum wire is very interesting because
specific properties can be easily achieved by varying the wire radius. An
electron bound to an impurity on the axis of the quantum wire behaves like a
bounded three-dimensional electron when the boundary is far away. However,
as the wire radius is reduced, the electron confinement due to the potential
barrier becomes very important. Especially in the quantum wire with
infinitely high potential wall, the total energy of the electron may change
from negative to positive at a certain radius and finally diverges to
infinity as the radius approaches zero. Furthermore, it is well known that
the reduction of dimensionality increases the effective strength of the
Coulomb interaction. The binding energy $E_{b}$ of the ground state of a
hydrogenic impurity in N-dimension is given by $E_{b}$=$\left[ \frac{2}{N-1}%
\right] ^{2}R_{y}^{*}$, where $R_{y}^{*}=\frac{\mu e^{4}}{2\varepsilon
^{2}\hbar ^{2}}$ is the effective Rydberg. Hence the dramatic change in the
binding energy may serve as a clear signal for variation in the effective
dimension of the quantum wire. \newline

It is known an electron weakly bound to a hydrogen impurity in a polar
semiconductor will interact with the phonons of the host semiconductor. In
the past decade, many authors have studied the polaron effect on the binding
energy of impurity or exciton in quantum well\cite%
{16,17,18,19,20,21,22,23,24}. Recently, the electron-phonon effect on the
binding energy of the donor impurity in a quantum wire with rectangular
cross-section was reported\cite{25,26,27}. It was found the polaron effect
on the binding energy becomes sizeable as the electron gets more deeply
bound. The polaron shifts in donor energy levels are found to be of the
order of 10\% in a weakly polar system. In studying the polaron effect on
the impurity binding energy, most of the previous works considered the
interaction of the electron and bulk optical(BO) phonon only. However, in
ionic crystal, the motion of an electron near the surface may be affected
very much by the surface longitudinal optical (SO) phonon\cite{28}. An
electron may be trapped at the surface by the electron-SO phonon
interaction. Besides, the electron phonon interaction Hamiltonian in the
previous works was valid only for the bulk. Therefore, we will choose the
Hamiltonian derived by Li and Chen\cite{29}, who considered the confined
phonon modes in the cylindrical quantum dot.

Most of the previous approaches concentrating on the polaron effect on the
ground state of an impurity in a quasi-one-dimensional wire employ the
variational method or perturbation method. Since the construction of
variational trial wave functions bases entirely on physical intuition, and
the estimation of the accuracy of the result obtained from variational
approach is very difficult. Furthermore, the perturbation method is only a
good access to those systems with very small perturbation in most cases.
Therefore, it would be most desirable to have an alternative approach which
is not only simple but also efficient to the quantum wire problem. In this
work, we employ a simple approximation treatment which combines the spirit
of both variational principle and perturbational approach to study the
effect of electron-phonon interactions on the ground state binding energy of
a hydrogenic impurity located inside a quantum wire.

\section{Theory}

\label{sec:theo}

Consider now a hydrogenic impurity located on the axis of a rigid wall
cylindrical quantum wire with a radius $d$. The Hamiltonian of the impurity
electron interacting with the phonon can be expressed as:\newline
\begin{equation}
H=H_{e}+H_{b}+H_{e-b}+H_{sp}+H_{e-sp},
\end{equation}
here, $H_{e}$ is the electronic part of the Hamiltonian 
\begin{equation}
H_{e}=-\frac{\hbar ^{2}}{2\mu }\left( \frac{\partial ^{2}}{\partial x^{2}}+%
\frac{\partial ^{2}}{\partial y^{2}}\right) -\frac{\hbar ^{2}}{2\mu }\frac{%
\partial ^{2}}{\partial z^{2}}-\frac{e^{2}}{\varepsilon r}+V(\rho )
\end{equation}
$V(\rho )$ is the confining potential which is assumed as: 
\begin{equation}
V(\rho )=\left\{ 
\begin{array}{l}
0\;\;\mbox{for}\;\rho \leq d, \\ 
\infty \;\;\mbox{for}\;\rho >d,%
\end{array}
\right.
\end{equation}
and $\varepsilon $ and $\mu $ are the dielectric constant of the well and
the effective mass of the electron. Recently, Li and Chen\cite{29} has
derived the confined the longitudinal-optical phonon and surface phonon
modes of a free-standing cylindrical quantum dot of radius $d$ and height $%
2D $. We will follow their Hamiltonian and let $D$ approach infinity, such
that the dot system can become a quantum wire. Therefore, $H_{b}$ is the
bulk phonon Hamiltonian which can be expressed as:

\begin{equation}
H_{b}=\sum_{n,l}\hbar \omega _{LO}a_{nl}^{\dag }a_{nl},
\end{equation}
where $\hbar \omega _{LO}$ is the dispersionless bulk optical (BO) phonon
energy, $a_{nl}^{\dag }(a_{nl})$ is the creation (annihilation) operator for
BO phonon. $H_{e-b}$ is the interaction between the electron and BO phonon
which can be expressed as:

\begin{eqnarray}
H_{e-b} &=&\sum_{n}J_{0}(\frac{\chi _{n}}{d}\rho
)[\sum_{l=1,3,...}V_{nl}\cos (\frac{l\pi }{2D}z)(a_{nl}+a_{nl}^{\dag }) 
\notag \\
&&+\sum_{l=2,4,...}V_{nl}\sin (\frac{l\pi }{2D}z)(a_{nl}+a_{nl}^{\dag })]
\end{eqnarray}%
with

\begin{equation}
V_{nl}=\frac{1}{V}\frac{4\pi e^{2}\hbar \omega _{LO}}{\left[ \left( \chi
_{n}/d\right) ^{2}J_{2}^{2}\left( \chi _{n}\right) +\left( l\pi /2D\right)
^{2}J_{1}^{2}(\chi _{n})\right] }\left( \frac{1}{\varepsilon _{\infty }}-%
\frac{1}{\varepsilon _{0}}\right) ,
\end{equation}
where $J_{m}$ is the mth-order Bessel function , $\chi _{n}$ is the nth-root
of $J_{0}$ , and\ $V=2\pi d^{2}D(D\longrightarrow \infty )$ is the crystal
volume. $H_{sp}$ is the surface optical phonon (SO) phonon Hamiltonian which
can be expressed as:

\begin{equation}
H_{sp}=\sum_{n}\hbar \omega _{sp}B_{n}^{\dag }B_{n},
\end{equation}
where $\hbar \omega _{sp}$ is the surface optical (SO) phonon energy, $%
B_{n}^{\dag }(B_{n})$ is the creation (annihilation) operator for SO phonon. 
$H_{e-sp}$ is the interaction between electron and SO phonon:

\begin{equation}
H_{e-sp}=\sum_{n=2,4,...}\Gamma _{n}I_{0}\left( \frac{n\pi }{2D}\rho \right)
\cos \left( \frac{n\pi }{2D}z\right) \left( B_{n}^{\dag }+B_{n}\right)
\end{equation}
with

\begin{eqnarray}
\Gamma _{n}^{2} &=&\frac{1}{S}\frac{4\pi e^{2}\hbar \omega _{sp}}{Dk_{n}%
\left[ I_{0}^{2}(k_{n}d)-I_{2}(k_{n}d)I_{0}(k_{n}d)\right] }  \notag \\
&&\cdot \left( \frac{1}{\varepsilon \left( \omega _{sp}\right) -\varepsilon
_{0}}-\frac{1}{\varepsilon \left( \omega _{sp}\right) -\varepsilon _{\infty }%
}\right) ,
\end{eqnarray}

\begin{equation}
\omega _{sp}^{2}=\left[ 1+\frac{\varepsilon _{0}-\varepsilon _{\infty }}{%
\varepsilon _{\infty }-\varepsilon \left( \omega _{sp}\right) }\right] ,
\end{equation}

\begin{equation}
\varepsilon \left( \omega _{sp}\right) =\frac{-I_{0}(k_{n}d)K_{1}(k_{n}d)}{%
K_{0}(k_{n}d)I_{1}(k_{n}d)},
\end{equation}
where $k_{n}=\frac{n\pi }{2D}$ , and $S=\pi d^{2}$ . $I_{m}$ and $k_{m}$ are
, respectively , the mth-order modified Bessel function of the first and
second kind.

Following Landau and Pekar's variational approach\cite{30}, the trial
wavefunction can be written as: 
\begin{equation}
\mid \Psi >\ =\ \Phi (r)U_{b}U_{s}\mid 0>,
\end{equation}
where $\Phi (r)$ depends only on the electron coordinate, and $\mid 0>$ is
the phonon vacuum state defined by $b_{q}\mid 0>=0$, $a_{q}\mid 0>=0$, and U
is a unitary transformation given by: 
\begin{eqnarray}
U_{b} &=&exp\left( \sum_{nl}(a_{nl}^{\dag
}\,f_{nl}-a_{nl}\,f_{nl}^{*})\right) , \\
U_{s} &=&exp\left( \sum_{n}(B_{n}^{\dag }\,g_{n}-B\,g_{n}^{*})\right) .
\end{eqnarray}
\noindent Where $f_{nl}$ and $g_{n}$ are the variational function and the
unitary operators $U_{b}$ and $U_{s}$ transform the bulk phonon and surface
phonon operators as follows: 
\begin{eqnarray}
U_{b}^{\dag }a_{nl}^{\dag }U_{b} &=&a_{nl}^{\dag }+f_{nl}^{\dag }, \\
U_{b}^{\dag }a_{nl}U_{b} &=&a_{nl}+f_{nl}, \\
U_{s}^{\dag }B_{n}^{\dag }U_{s} &=&B_{n}^{\dag }+g_{n}^{\dag }, \\
U_{s}^{\dag }B_{n}U_{s} &=&B_{n}+g_{n}.
\end{eqnarray}
The parameters $f_{q}$, $f_{q}^{*}$, $g_{q}$, $g_{q}^{*}$ can be obtained by
minimizing the $<\mid H\mid >$ with respect to the parameters $f_{nl}$, $%
f_{nl}^{*}$, $g_{n}$, $g_{n}^{*}$. Then the expectation value $<H>$ turns
out to be

\begin{eqnarray}
&<&\ H\ >=<\Phi (r)\mid H_{e}\mid \Phi (r)> \\
&&-\sum_{nl}\frac{V_{nl}^{2}}{\hbar \omega _{LO}}\left| <\Phi (r)\mid J_{0}(%
\frac{\chi _{n}}{d}\rho )\cos (\frac{n\pi }{2D}z)\mid \Phi (r)>\right| ^{2} 
\notag \\
&&\qquad -\sum_{n}\frac{\Gamma _{n}^{2}}{\hbar \omega _{sp}}\left| <\Phi
(r)\mid I_{0}(k_{n}\rho )\cos (\frac{n\pi }{2D}z)\mid \Phi (r)>\right| ^{2} 
\notag
\end{eqnarray}

The axis of the wire is assumed to be along the z direction. To solve the
electronic part, one can employ the perturbative-variational approach as we
did in the above subsection. Two variational parameters $\alpha $ and $\beta 
$ are introduced by adding and subtracting two terms $\frac{\alpha e^{2}}{%
\varepsilon \rho }$ and $\frac{\beta ^{2}\hbar ^{2}}{2\mu }z^{2}$ into the
original Hamiltonian $H_{e}$ and then regroup $H_{e}$ into three groups:%
\newline
\begin{equation}
H_{e}=H_{01}(\beta )+H_{02}(\alpha )+H^{\prime }(\alpha ,\beta )
\end{equation}
\noindent where 
\begin{eqnarray}
H_{01}(\beta ) &=&\frac{-\hbar ^{2}}{2\mu }\frac{\partial ^{2}}{\partial {z}%
^{2}}+\frac{\beta ^{2}\hbar ^{2}}{2\mu }z^{2}, \\
H_{02}(\alpha ) &=&\frac{-\hbar ^{2}}{2\mu }(\frac{\partial ^{2}}{\partial
x^{2}}+\frac{\partial ^{2}}{\partial y^{2}})-\frac{\alpha e^{2}}{\varepsilon
\rho }+V(\rho ) \\
H^{\prime }(\alpha ,\beta ) &=&\frac{\alpha e^{2}}{\varepsilon \rho }-\frac{%
\beta ^{2}\hbar ^{2}}{2\mu }z^{2}-\frac{e^{2}}{\varepsilon r}.
\end{eqnarray}
\noindent In the above equations, $H^{\prime }(\alpha ,\beta )$ is treated
as a perturbation, and $\alpha $ and $\beta $ are treated as variational
parameters which can be determined by requiring the perturbation term to be
as small as possible. Decomposing $H_{e}$ into two terms $H_{01}$ and $%
H_{02} $ is equivalent to dividing the space into a two-dimensional (in xy
plane) and a one-dimensional (in z-axis) subspace. The unperturbed part of
the Hamiltonian $H_{e}$ contains two terms, i.e. $H_{01}$ and $H_{02}$,
where $H_{01}$ represents the one dimensional harmonic oscillator, and $%
H_{02}$ represents a two dimensional hydrogen atom located inside a quantum
disk\cite{14}. Both can be solved exactly. For illustration, the ground
state energy and wavefunction of the unperturbed part can be expressed as : 
\begin{eqnarray}
E_{g}^{(01)}(\alpha ,\beta ) &=&E_{g}^{(01)}(\beta )+E_{g}^{(02)}(\alpha ) \\
\Psi _{g}^{(0)}(r,\alpha ,\beta ) &=&\varphi _{g}^{(01)}(z;\beta )\varphi
_{g}^{(02)}(x,y;\alpha )
\end{eqnarray}
respectively, where $\varphi _{g}^{(01)}(z;\beta )$ is the ground state
wavefunction of the 1D harmonic oscillator, and $\varphi
_{g}^{(02)}(x,y;\alpha )$ is the ground state wavefunction of the 2D
hydrogen atom located at the center of an infinite circular well. The ground
state eigenvalue and eigenfunction of the 1D harmonic oscillator can be
expressed as: 
\begin{eqnarray}
E_{g}^{(01)}(\beta ) &=&\frac{\beta \hbar ^{2}}{2\mu } \\
\varphi _{g}^{(01)}(z;\beta ) &=&(\frac{\beta }{\pi })^{\frac{1}{4}}e^{-%
\frac{\beta }{2}z^{2}}
\end{eqnarray}
The ground state eigenvalue and eigenfunction of the 2D hydrogenic impurity
located at the center of an infinite circular well can be obtained as\cite%
{14}:\newline
(1) For $E<0$, 
\begin{equation}
\varphi _{g}^{(02)}(x,y;\alpha )=N_{1}e^{-\frac{\xi _{1}}{2}}\xi _{1}^{\mid
\,m\,\mid \;+1}\Phi (\mid \,m\,\mid +1/2-\lambda _{1},2\mid \,m\,\mid +1,\xi
_{1})
\end{equation}
where $\xi _{1}=\alpha _{1}\rho $, $\alpha _{1}=\frac{-8\mu E}{\hbar ^{2}}$, 
$\lambda _{1}=\frac{2\mu \alpha e^{2}}{\varepsilon \hbar ^{2}\alpha _{1}}$, $%
\Phi (a,b,x)$ is the confluent hypergeometric function, and $N_{1}$ is the
normalization constant.\newline
(2) For $E>0$, 
\begin{equation}
\varphi _{g}^{(02)}(x,y;\alpha )=N_{2}\xi _{2}^{m}\Phi _{m-\frac{1}{2}}(\eta
_{2},\xi _{2})
\end{equation}
where $\xi _{2}=\alpha _{2}\rho $, $\alpha _{2}=\frac{-8\mu E}{\hbar ^{2}}$, 
$\eta _{2}=\frac{-\mu \alpha e^{2}}{\varepsilon \hbar ^{2}\alpha ^{2}}$, $%
\Phi _{m-1/2}(\eta _{2},\xi _{2})$ is the irregular Coulomb wave function,
and $N_{2}$ is the normalization constant.\newline
(3) The turning point for energy changing from $E>0$ to $E<0$ in the quantum
circle system may be determined by setting 
\begin{equation}
d^{-1/2}J_{0}\left[ \left( \frac{8\mu e^{2}}{\varepsilon \hbar ^{2}}\right)
^{1/2}d^{1/2}\right] =0\;\;for\;\;m=0,
\end{equation}
and\newline
\begin{equation}
d^{-1/2}J_{2}\left[ \left( \frac{8\mu e^{2}}{\varepsilon \hbar ^{2}}\right)
^{1/2}d^{1/2}\right] =0\;\;for\;\;m=1,
\end{equation}
The requirement of the continuity of the wavefunctions and its first
derivative at boundary yields:\newline
(1) For $E<0$, 
\begin{equation}
\phi (\mid \,m\,\mid \;+\frac{1}{2}-\lambda _{1},2\mid \,m\,\mid \;+1,\alpha
,d)=0
\end{equation}
(2) For $E>0$ 
\begin{equation}
\Phi _{m-1/2}(\eta _{2},\alpha _{2}d)=0
\end{equation}
The eigenvalues are then given as: 
\begin{equation}
E_{g}^{(02)}(\alpha )=\left\{ 
\begin{array}{l}
-\frac{\mu \alpha ^{2}e^{4}}{2\varepsilon ^{2}\hbar ^{2}\lambda _{1}^{2}}\;\;%
\mbox,\;for\;E<0, \\ 
\frac{\mu \alpha ^{2}e^{4}}{2\varepsilon ^{2}\hbar ^{2}\eta _{2}^{2}}\;\;%
\mbox,\;for\;E>0,%
\end{array}
\right.
\end{equation}
The first order energy correction can thus obtained as: 
\begin{eqnarray*}
\Delta E_{g}^{(1)}(\alpha ,\beta ) &=&<\Phi _{g}^{(0)}(r;\alpha ,\beta )\mid
H^{\prime }(\alpha ,\beta )\mid \Phi _{g}^{(0)}(r;\alpha ,\beta )> \\
&=&\mbox{}<\varphi _{g}^{(02)}(x,y:\alpha )\mid \frac{\alpha e^{2}}{%
\varepsilon \rho }\mid \varphi _{g}^{(02)}(x,y;\alpha )> \\
&&-<\varphi _{g}^{(01)}(z;\beta )\mid \frac{\beta ^{2}\hbar ^{2}z^{2}}{2\mu }%
\mid \varphi _{g}^{(01)}(z;\beta )> \\
&&-<\Phi _{g}^{(0)}(r;\alpha ,\beta )\mid \frac{e^{2}}{\varepsilon r}\mid
\Phi _{g}^{(0)}(r;\alpha ,\beta )>
\end{eqnarray*}
The second term of the above equation can be integrated analytically and the
result is: \newline
\begin{equation}
<\varphi _{g}^{(01)}(z;\beta )\mid \frac{\beta ^{2}\hbar ^{2}z^{2}}{2\mu }%
\mid \varphi _{g}^{(01)}(z;\beta )>\varphi _{g}^{(02)}(x,y;\alpha )>=\frac{%
\beta \hbar ^{2}}{\mu }.
\end{equation}
Then the total energy up to the first order perturbation correction can then
be obtained as: 
\begin{equation}
E_{g}(\alpha ,\beta )=E_{g}^{(01)}(\beta )+E_{g}^{(02)}(\beta )+\Delta
E_{g}^{(1)}(\alpha ,\beta ),
\end{equation}
The variational parameters are then chosen by requiring the total energy $%
E_{g}(\alpha ,\beta )$ to be minimized with respect to the variation of $%
\alpha $ and $\beta $. This is equivalent to requiring: 
\begin{equation}
\frac{\partial E}{\partial \alpha }=0,
\end{equation}
\begin{equation}
\frac{\partial E}{\partial \beta }=0.
\end{equation}
For the excited states, the eigenvalues and eigenfunctions can be treated in
the same way.\newline

\section{Results and Discussions}

We have calculated the effect of the confined the longitudinal-optical
phonon and surface phonon interactions on the hydrogenic impurity located in
a quantum wire. And the well potential is considered as infinite. 
\begin{figure}[h]
\includegraphics[width=8cm]{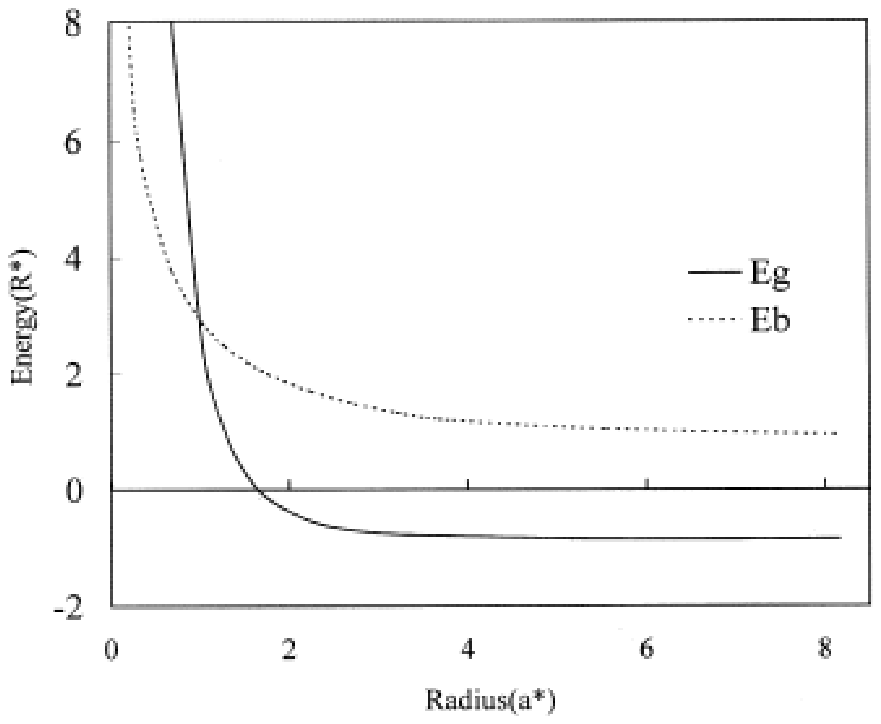}
\caption{The ground-state energy(solid line) and the binging energy (dotted
line) of a hydrogenic impurity located at the axis of a cylindrical wire as
a function of the radius of the wire. (Ry* and a* are effective rydberg and
the effective Bohr radius.)}
\end{figure}
Figure 1 shows the ground state energy as a function of the wire radius. The
binding energy $E_{b}$ of the hydrogenic impurity is defined as the energy
difference between the ground state energy of the cylindrical wire system
with and without the impurity, i.e.\newline
\begin{equation}
E_{b}=E_{0}-E_{g}
\end{equation}%
where $E_{0}$ is the ground state energy of the quantum wire system without
the impurity, while $E_{g}$ is the ground state energy of the quantum wire
system with the impurity located on the axis of the cylindrical wire. One
can see from Fig.1 that the energy of the 1$s$ state becomes negative when
the wire radius is larger than 1.65$a^{\ast }$. It means that the confining
energy is larger than the Coulomb energy as the wire radius is smaller than
1.65$a^{\ast }$. And one can also note that as the radius of the quantum
wire is decreased, the ground state energy increases. As the wire radius $d$%
\ becomes smaller, the electron is pushed toward the axis of the cylindrical
wire. This makes the electron get close to the nucleus. As the electron gets
close to the nucleus, both the ground state energy and the binding energy
increase rapidly. This is because the Coulomb potential, which varies with $%
\sim \frac{1}{d}$\ ($d$\ is the wire radius), becomes more negative, while
the kinetic energy of the electron, which varies with $\sim \frac{1}{d^{2}}$%
\ (by the uncertainty relation), increases more rapidly. As a result, the
ground state energy is increased as the electron gets close to the nucleus.
The binding energy defined in Eq. (39) is effectively the negative sign of
the of the Coulomb interaction energy between the electron and the nucleus,
i.e. $\sim \frac{1}{d}$, therefore, the binding energy of the electron is
also increased as the electron gets near to the nucleus.. As a result, the
ground state energy is increased as the electron gets close to the nucleus.
The binding energy defined in Eq. (39) is effectively the negative sign of
the of the Coulomb interaction energy between the electron and the nucleus,
i.e. $\sim \frac{1}{d}$, therefore, the binding energy of the electron is
also increased as the electron gets near to the nucleus. Our results show
that for small wire radius, the binding energies are in good agreement with
previous results\cite{11,14}. As the radius becomes very large, our result
approaches the correct limit 1$R^{\ast }$ while the previous work \cite{14}
can only yield a value of 0.22$R^{\ast }$. The large discrepancy of the
previous work may be due to the artificial dividing of the variational trial
wavefunction into a one-dimensional hydrogen atom and a two-dimensional
hydrogen atom and thus forces the creation of an additional node of the
wavefunction at z=0. In this work, the trial wavefunction is adopted to be
the form of 1D harmonic oscillator wavefunction instead of one dimensional
hydrogen atom. This prevents our wavefunction from introducing any
additional node at z=0.Figure 2 presents the 2s excited state binding
energies as the functions of wire radius. One can note from the figure that
as the wire radius increases, the binding energy approaches 0.25$R^{\ast }$
which gives correctly the limiting value of 3D hydrogen atom. 
\begin{figure}[h]
\includegraphics[width=8cm]{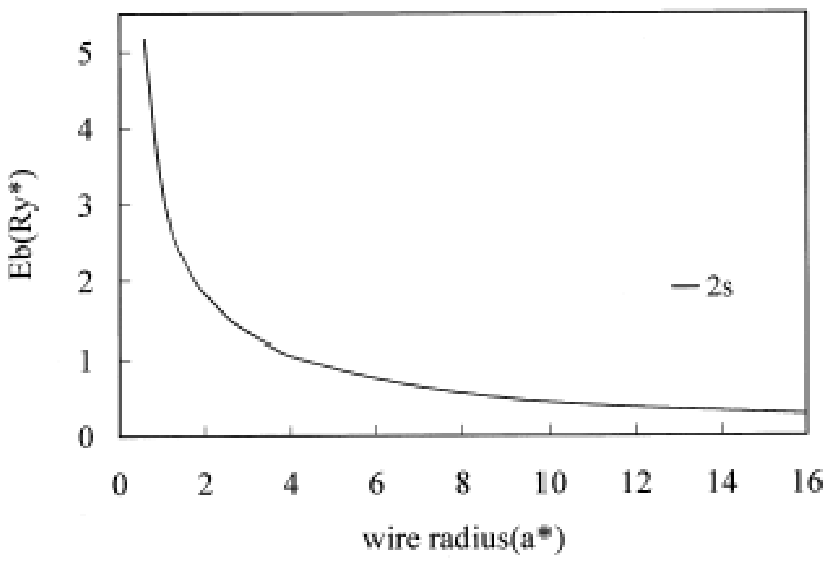}
\caption{The binding energy of the 2s excited state as a function of the
wire radius. (Ry* and a* are effective rydberg and the effective Bohr
radius.)}
\end{figure}
\begin{figure}[h]
\includegraphics[width=8cm]{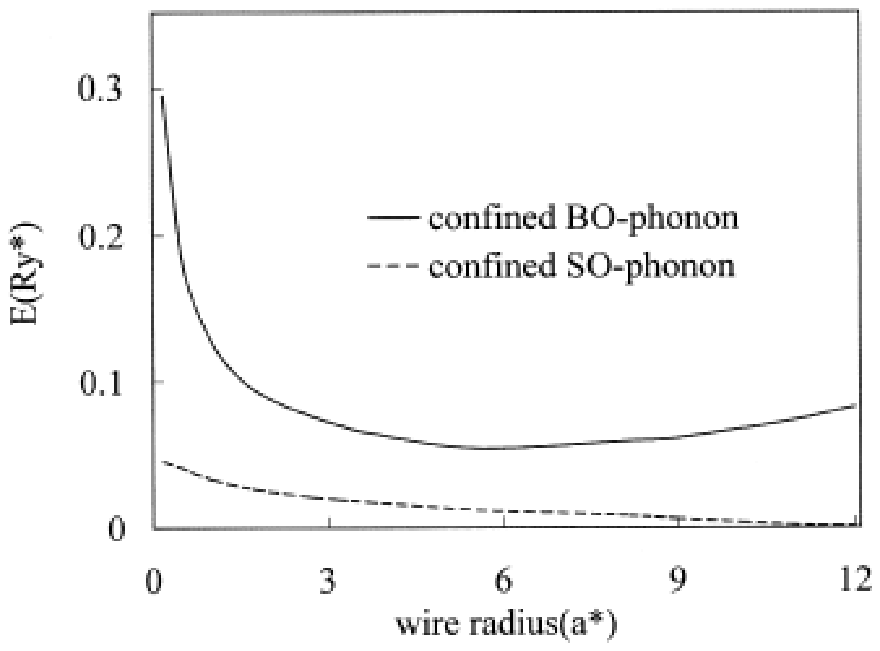}
\caption{The energies of the wire radius modified by the confined BO and SO
phonon. The solid line stands for the BO phonon effect, and the dashed line
for the SO phonon effect. (Ry* and a* are effective rydberg and the
effective Bohr radius.)}
\end{figure}

Figure 3 presents the confined BO phonon and SO phonon effect as a function
of wire radius. With increasing the wire radius, the magnitude of the
confined BO phonon effect decreases from large value and then approaches to
the bulk value. When the wire radius is less than 1.5 a*, the polaron effect
increases rapidly. One might think as the radius becomes very small, the
confined BO phonon effect should approach zero, like the case in quantum
well \cite{31}. In fact, similar results were obtained by Oshiro in a
spherical quantum dot\cite{32}. They found the polaron energy shift is
enhanced as the dot radius becomes small. This is due to the fact that the
electron becomes complete localized (E$_{b}$ approaches infinity) in small
wire (or dot) radius while the binding energy approaches 4$R^{\ast }$ in
small well width. In the case of quantum well, the confined SO phonon effect
plays the dominant role for small well width\cite{31}. But in quantum wire,
the confined SO phonon is less important, just like that in quantum dot
system \cite{32}. This is because the surface area of a quantum wire (or
quantum dot) decreases with the radius. Thus the number of vibration modes
of confined SO phonon becomes fewer. 
\begin{figure}[h]
\includegraphics[width=8cm]{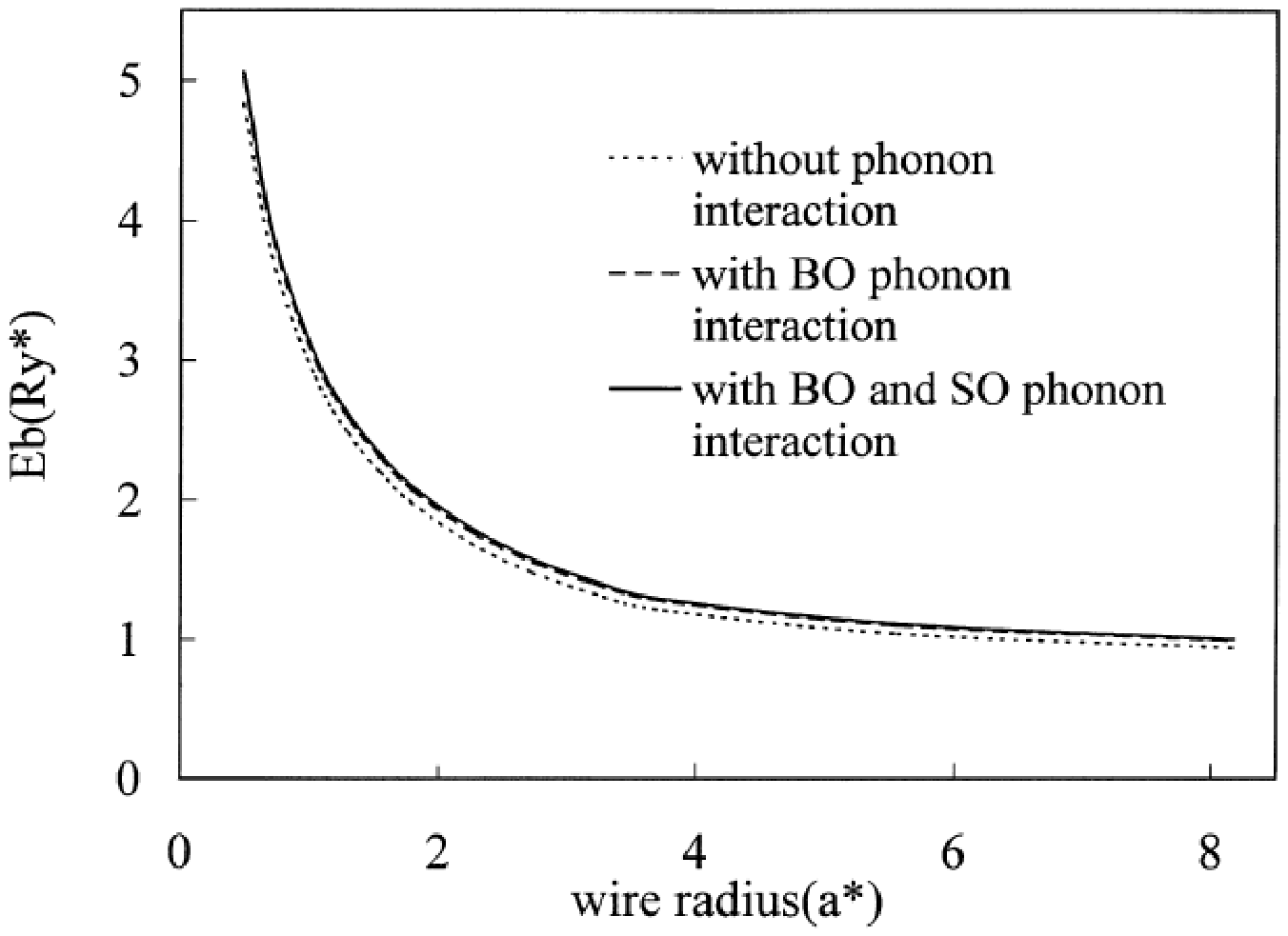}
\caption{The binding energy with/without phonon effect. The dotted line
stands for the binding energy without the phonon effect. The dashed line
stands for that only BO phonon effect on the binding energy, and the solid
line for both the BO and SO phonons effects on the binding energy. (Ry* and
a* are effective rydberg and the effective Bohr radius.)}
\end{figure}

In Fig.4, three curves are presented. The dotted curve represents the
binding energy of the impurity without considering the interactions between
the electron and phonon. The dashed curve represents the binding energy of
the impurity with only confined BO phonon effect being taken into account.
While the solid curve is the binding energy of the impurity including both
confined BO phonon and SO phonon effects in the calculation. Comparing to
the impurity binding energy, the confined SO phonon is negligible in quantum
wire. We then conclude that because of the similarity in geometry, the
behavior of the polaron effect on the quantum wire system is like that on
the quantum dot system.

\section{Conclusion}

In this work, analytical solutions for the effects of the electron-phonon
interaction on the binding energies of an impurity located inside a quantum
wire are obtained by a simple but efficient perturbation-variation method.
As the radius becomes very large, the correct limiting value can be
obtained. We have also discussed both the confined BO and SO phonon effects.
We found the confined BO phonon effect is prominently for a quantum wire
with small radius. We also found that the energy corrections of the polaron
effects on the impurity binding energies increase rapidly when the wire
radius is less than 1.5 a*.

\noindent \textbf{Acknowledgements}

This work is supported partially under the grant number NSC
88-2112-M-009-004 by the National Science Council, Taiwan.\newline

\end{document}